\documentclass[11pt]{article} % use larger type; default would be 10pt

\usepackage[utf8]{inputenc} % set input encoding (not needed with XeLaTeX)

\usepackage[margin=1in]{geometry}% to change the page dimensions
\usepackage{graphicx} % support the \includegraphics command and options
\graphicspath{ {./Figures/} }

\usepackage{booktabs} % for much better looking tables
\usepackage{array} % for better arrays (eg matrices) in maths
\usepackage{paralist} % very flexible & customisable lists (eg. enumerate/itemize, etc.)
\usepackage{verbatim} % adds environment for commenting out blocks of text & for better verbatim
\usepackage{caption}
\usepackage{subcaption}
\usepackage{xcolor} %Colors
\usepackage{mathrsfs} %

\usepackage{fancyhdr} % This should be set AFTER setting up the page geometry
\usepackage{sectsty}
\allsectionsfont{\sffamily\mdseries\upshape} % (See the fntguide.pdf for font help)
\usepackage[nottoc,notlof,notlot]{tocbibind} % Put the bibliography in the ToC
\usepackage[titles,subfigure]{tocloft} % Alter the style of the Table of Contents

\usepackage[hidelinks]{hyperref}
\usepackage{amsmath,amsthm,mathtools}
\usepackage{amssymb}

\title{ Seeking Dark Signals in Oscillating Redshifts: Exploring Geometric Scalar Field Dark Matter}

\begin{document}

\author{James Wheeler}
\date{}
\maketitle

\begin{abstract}
We detail a novel theoretical prediction that a geometric torsion model for scalar field dark matter could lead to oscillations, on readily probeable timescales, in the time evolution of cosmological redshifts of astronomical sources with qualitatively distinct behavior at different redshift scales (larger or smaller than $z \sim 0.1$). We present an analysis of extant spectroscopy data from the Australian Dark Energy Survey (OzDES) to assess whether such signals are present across a wide array of cosmological sources and baseline redshifts on the six-year timescale of OzDES. While a simple Fourier analysis of redshift variations weakly identifies some candidate frequencies, and so further investigation with future cosmological data sets may be warranted, we have not found compelling empirical evidence for the theory under consideration in this data set, placing tentative constraints on its free parameters.
\end{abstract}

\section{Introduction}

The problem of identifying the nature of dark matter remains one of the largest challenges facing theoretical physics today-- evidence abounds for a consistent discrepancy between the observed gravitational dynamics of the universe, across all scales much larger than the solar system, and what is expected based upon the amount of baryonic matter detected or inferred, generally pointing to a matter deficit. A class of models which has received much attention are broadly categorized as Scalar Field Dark Matter (SFDM; other relevant monikers include ``fuzzy" or ``wave" dark matter), characterized by dark matter's being classically well-modeled by a scalar field $\phi$ coupled to the Einstein equation, e.g.\@ as in equation (\ref{eqn:einsteinkg}) below \cite{hu2000fuzzy, matos2009varphi2}. SFDM is often considered with an extremely light mass parameter $m \sim 10^{-22}$ eV since this naturally suppresses structure formation on small scales (with an absolute minimum set by Compton length scale $\hbar c / m$), though many observational constraints favor larger masses from a variety of sources  \cite{armengaud2017constraining,chiang2021soliton,davoudiasl2019ultralight,hlozek2015search,schutz2020subhalo}: these tend to roughly prefer $m \gtrsim 10^{-21}$ eV, though a more recent work extends as far as $m \geq 3 \cdot 10^{-19}$ eV \cite{dalal2022excluding}. The most common means of fundamentally motivating SFDM is the invocation of an ultralight axion \cite{chadha2022axion,huiwitten,marsh2016axion}, though for many purposes SFDM can be investigated purely phenomenologically with only (\ref{eqn:einsteinkg}) since dark matter is empirically a classical phenomenon. 

In 2010, Bray \cite{bray2010} showed that one can also motivate SFDM entirely classically via a natural modification to the geometry of general relativity, namely by allowing a nontrivial connection to contribute to the action. In this work, we investigate some potential implications of this geometric adjustment when one treats the nontrivial connection as having physical content beyond its implications for the Euler-Lagrange equations. We find that treating Bray's axioms broadly can lead to a rather distinct prediction for the behavior of gravitational redshifts in the presence of variations of the scalar field $\phi$, encapsulated in equation (\ref{eqn:redshift}). In particular, this could have readily evaluable implications for the time evolution of cosmological redshifts, quantities which, as one of the primary observables by which we probe and characterize the universe, have undergone a large degree of empirical scrutiny.

This scrutiny notwithstanding, the time evolution of redshifts of fixed sources has not been thoroughly investigated empirically. This is largely due to the fact that the expected rate of change, under the standard model of cosmology, of the redshift of a source at fixed comoving distance is comparable (for redshifts $z \leq 10$) to the Hubble parameter in order of magnitude,
\begin{equation} \label{eqn:zdot}
H = h \cdot  \frac{100 \text{ km/s}}{\text{Mpc}} \approx h \cdot \frac{10^{-10}}{\text{yr}}.
\end{equation}
Though theorists have toyed with the idea of detecting this for at least half a century \cite{sandage1962change}, it has remained hopelessly outside the reach of direct measurement on reasonable timescales, perhaps until very recently \cite{balbi2007time, lake2007testing, loeb1998direct}. Indeed, it is only recently that surveys have begun collecting high-quality spectroscopy data for fixed sources repeatedly over many years, though this has largely been done with an interest in the reverberation mapping of Active Galactic Nuclei (AGNs) rather than redshift evolution \cite{peterson1993reverberation, swann20194most, yuan2015ozdes}. We utilize the catalogued data of one such recently-completed survey, the Australian Dark Energy Survery (OzDES) \cite{lidman2020ozdes, yuan2015ozdes}, to investigated the time evolution in the redshifts of 1457 distinct sources in order to assess the empirical standing of the predictions of the geometric model of SFDM considered herein, ultimately placing constraints on the parameters of the theory in equation (\ref{eqn:constraints}).

This paper is organized as follows. The following section presents and discusses the theory under consideration, deriving the general redshift adjustment as well as its specialization to the cosmological context. Section \ref{sec:methods} describes the data set under scrutiny and formulates the computational problem of, and approach to, extracting redshifts from the catalogued spectroscopy data. Section \ref{sec:results} presents the results of our analysis, largely contained in Figures \ref{fig:redshiftvar}-\ref{fig:dft}, and the constraints derived on the theory. Section \ref{sec:conclusions} reflects on the work and puts forward some concluding remarks. The appendix furnishes additional detail on the formulation and derivation of the theory from a modified action. All work is done in natural units ($\hbar = c = 1$).

\section{Theory} \label{sec:theory}

\subsection{A Geometric Picture of Scalar Field Dark Matter}
The Einstein-Hilbert action $S$ of general relativity, on a smooth manifold $M$, is a functional of a Lorentzian metric $g$ on $M$ given by
$$ S[g] = \int_U R \; dV,$$
where $R$ and $dV$ are (respectively) the scalar curvature and volume form associated to $g$ and $U$ is any open set with compact closure in $M$. This action characterizes vacuum general relativity (without cosmological constant) in that requiring $g$ to a be a critical point of $S$ for every $U$ is equivalent to the vacuum Einstein equation $G = 0$. This provides a particularly compelling picture due to a result of Lovelock \cite{lovelock1971einstein} indicating that $S$ is, in fact, the unique coordinate-invariant action quadratic in the derivatives of $g$. Bray \cite{bray2010} demonstrated in 2010 that minimally relaxing this constraint to allow a general connection $\nabla$ of the spacetime manifold-- by allowing $S[g,\nabla]$ to be quadratic in $g_{ij,k}$ and the connection coefficients $\Gamma_{ijk}$ and their derivatives $\Gamma_{ijk,l}$-- generically leads to the inclusion of a massive scalar field source term in the Einstein equation, resulting in the Einstein-Klein-Gordon system with cosmological constant:
\begin{equation} \label{eqn:einsteinkg}
G + \Lambda g = 8 \pi G \left[ 2 \frac{d \phi \otimes d \phi}{m^2} - \left( \frac{ | d \phi | ^2}{m^2} + \phi^2 \right) g \right], \qquad \Box \phi = m^2 \phi
\end{equation}
for some scalar field $\phi : M \to \mathbb{R}$ intimately tied to the connection $\nabla$. This provides a geometric motivation for considering a scalar field as a potential dark matter candidate.

If one is to take this geometric picture seriously as a framework giving rise to dark matter, we should consider the question of what the connection $\nabla$ indicates physically: what is the physical distinction between this theory and one incorporating the Levi-Civita connection? The most natural hypothesis is that $\nabla$ provides the geodesics along which test particles and light propagate, according to the coordinate geodesic equation
\begin{equation} \label{eqn:geodesic}
\ddot \gamma^k + \Gamma_{ij}^{\;\;\, k} \dot \gamma^i \dot \gamma^j = 0,
\end{equation}
where $\Gamma_{ijk} = \langle \nabla_{\partial_i} \partial_j, \partial_k \rangle$ is the Christoffel symbol associated to $\nabla$. In the simplest nontrivial instantiation of a theory conforming to the relaxed geometric axiom put forward by Bray, and the primary scenario he presented, one finds that $\nabla$ is related to $g$ and $\phi$ according to

\begin{equation}\label{eqn:oldchristoffel}
\Gamma_{ijk} = \overline \Gamma_{ijk} + (*d \phi)_{ijk},
\end{equation}
where $\overline \Gamma$ is the Christoffel symbol of the Levi-Civita connection $\overline \nabla$ of $g$ and $*$ denotes the Hodge star operation. In particular, the difference between $\Gamma$ and $\overline \Gamma$ is entirely antisymmetric in this case, implying that $\nabla$ and $\overline \nabla$ have the same geodesics according to (\ref{eqn:geodesic}). That is, this simplest case accommodates standard general relativity with its only primary modification being the addition of the scalar field source (though more general parallel propagation would be adjusted). 

In the interest of exploring the range of physical phenomena this geometric picture might give rise to, here we would like to consider the next-simplest case beyond that described above. This involves still taking $\nabla$ to be metric compatible, as one would want physically if its geodesics are to describe test particles, but allowing the difference tensor $D_{ijk} := \Gamma_{ijk} - \overline \Gamma_{ijk}$ to have a nontrivial trace form $ D^{\;\;\, j}_{ji}$. The equations of motion in this scenario still give the Einstein-Klein-Gordon system (\ref{eqn:einsteinkg}), but with the Christoffel symbol relation (\ref{eqn:oldchristoffel}) modified to

\begin{equation}\label{eqn:newchristoffel}
\Gamma_{ijk} = \overline \Gamma_{ijk} + (*d \phi)_{ijk} + C \left[ (d \phi)_j g_{ik} - (d \phi)_k g_{ij}\right],
\end{equation}
where $C$ is a free parameter of the theory, in the same vein as $m$. See Appendix \ref{app:theory} for a detailed derivation and discussion of this key formula. In less coordinate-laden parlance, this can be written
\begin{equation} \label{eqn:newchristoffel2}
\langle \nabla_{X}Y, Z \rangle =  \langle \overline \nabla_{X}Y, Z \rangle + (*d \phi)(X,Y,Z) + C \left[ Y( \phi ) \langle X, Z \rangle + Z( \phi ) \langle X,Y \rangle \right]
\end{equation}
for any vector fields $X,Y,Z$. 

\subsection{The General Redshift Adjustment}

We investigate the implications of (\ref{eqn:newchristoffel2}) for geodesics. When evaluating whether a given curve through spacetime is a geodesic, one is interested in $\nabla_T T$ with $T$ the tangent vector field to the curve, for which the hodge star term above is null by antisymmetry:

$$ \langle \nabla_{T}T, Z \rangle =  \langle \overline \nabla_{T}T, Z \rangle + C \left[ T( \phi ) \langle T, Z \rangle + Z( \phi )  | T |^2 \right].$$
Observing $Z( \phi ) = \langle \nabla \phi, Z \rangle$ (recall that $\nabla \phi = \overline \nabla \phi = \text{grad } \phi$ is constructed out of the metric $g$ independently of the connection), we notice that the entire righthand side may be written in the form $\langle \, \cdot \, , Z\rangle$, and so nondegeneracy of the metric allows us to deduce

$$\nabla_T T = \overline \nabla_T T + C \left[ T( \phi ) T+| T |^2 \nabla \phi \right].$$

Supposing that $T$ is the tangent vector field to a geodesic of the Levi-Civita Connection $\overline \nabla$ (so that $\overline \nabla_T T = 0$), then, we've found that 
$$\nabla_T T = C \left[ T( \phi ) T+| T |^2 \nabla \phi \right].$$
In general, this equation means that a geodesic of $\overline \nabla$ is no longer a geodesic of $\nabla$, since the righthand side is not universally $0$ so long as $C \nabla \phi \neq 0$, meaning that the {\it variation} of the dark matter scalar field $\phi$ can impact the trajectories of test particles beyond its usual gravitational influence mediated by the metric. Considering the particular case of a null geodesic to $\overline \nabla$ to understand implications for light, the result finally reduces to 
\begin{equation} \label{eqn:nullgeo}
\nabla_T T = C T( \phi ) T.
\end{equation}
In this case, that $\nabla_T T$ is parallel to $T$ means that the trajectory giving rise to $T$ is still that of a geodesic, but its geodesic {\it parameterization} has changed. This parameterization is what determines the gravitational redshift of light following the trajectory in question, leading us to the potential for an easily observable signal in redshifts. Let us compute the general adjustment to the gravitational redshift in this theory before specializing to the standard FLRW cosmology.

If $\gamma : I \to M$ (for some interval $I \subset \mathbb{R}$) is a null geodesic of $\overline \nabla$, we wish to compute how $\gamma$ should be reparameterized according to a reparameterizing function $s \mapsto \tau(s)$ to obtain a geodesic $\tilde \gamma(s) := \gamma (\tau(s))$ of $\nabla$. Then $\tilde \gamma '(s) = \tau'(s) \gamma '(\tau(s))$, and we find  
\begin{align*}
\nabla_{\tilde \gamma'(s)} \tilde \gamma'(s) & = \nabla_{\tilde \gamma'(s)} \left[ \tau'(s) \gamma'(\tau(s)) \right] 
\\ & = \left[ \nabla_{\tilde \gamma '(s)} \tau'(s) \right] \gamma'(\tau(s)) + \tau'(s) \left[ \nabla_{\tilde \gamma '(s)} \gamma '(\tau(s)) \right] 
\\ & = \tau''(s) \gamma'(\tau(s)) + (\tau'(s))^2 \nabla_{ \gamma'( \tau(s))} \gamma '(\tau(s))
\\ & = \left[ \tau''(s) + C (\tau'(s))^2  \gamma'(\tau(s))[\phi]  \right] \gamma'(\tau(s))
\\ & = \left[ \tau''(s) + C \tau'(s) (\phi \circ \tilde \gamma)'(s) \right] \gamma' (\tau(s)),
\end{align*}
where we have used (\ref{eqn:nullgeo}) to replace $\nabla_{ \gamma'( \tau(s))} \gamma '(\tau(s))$ as well as that the action of $\tilde \gamma'(s)$ on $\phi$ results in $(\phi \circ \tilde \gamma)'(s)$ by definition of the action of a tangent vector on a function. Hence, requiring that this be $0$ so that $\tilde \gamma$ is a geodesic of $\nabla$ leads us to an ODE for $\tau(s)$:

$$ \tau''(s) + C \tau'(s) (\phi \circ \tilde \gamma)'(s) = 0 .$$
The general solution satisfies
$$\tau'(s) = K e^{- C \phi(\tilde \gamma(s))},$$
with $K \in \mathbb{R}$ an arbitrary constant.

If observers at the points $p_1 = \tilde \gamma(s_1)$ and $p_2 = \tilde \gamma(s_2)$ following worldlines with tangent vectors $T_1$ and $T_2$ (in the cosmological case that follows, $T_1$ and $T_2$ are both the tangent vector field $\frac{\partial}{\partial t}$ to comoving observers) measure the frequency of a light ray propagating along $\tilde \gamma$, they measure frequencies proportional to $\tilde \omega_i = \langle T_i, \tilde \gamma'(s_i) \rangle$, meaning that between them they observe a redshift
\begin{align} \label{eqn:redshift}
1+ \tilde z = \frac{\tilde \omega_1}{\tilde \omega_2} & = \frac{\langle T_1, \tilde \gamma'(s_1) \rangle}{ \langle T_2, \tilde \gamma'(s_2) \rangle }   \nonumber
\\ & = \frac{\tau'(s_1)}{\tau'(s_2)} \cdot \frac{\langle T_1, \gamma'(\tau_1) \rangle}{ \langle T_2, \gamma'(\tau_2) \rangle }   \nonumber
\\ &= e^{C [ \phi(p_2) - \phi(p_1) ] } \cdot  \frac{\omega_1}{\omega_2}  \nonumber
\\ \Aboxed{ 1 + \tilde z &= (1+z) e^{C [ \phi(p_2) - \phi(p_1) ] }, } 
\end{align}
where quantities with a tilde correspond to light propagating along $\tilde \gamma$ (in accordance with $\nabla$) and quantities without a tilde correspond to light propagating along $\gamma$ (in accordance with $\overline \nabla$). Equation (\ref{eqn:redshift}) is finally the general adjustment to the gravitational redshift expected within this geometric framework for scalar field dark matter, assuming the next-to-simplest admissible action $S[g,\nabla]$ yielding metric compatibility (equation (\ref{eqn:action1})) and that the nontrivial connection $\nabla$ manifests physically in the trajectories of test particles. It indicates that the redshift expected under $\nabla$ is, in general, that expected under $\overline \nabla$ modulated by the change in the value of the scalar field $\phi$ between observation and emission, with the degree of modulation set by the free parameter $C$ of the theory (which evidently has units, under $\hbar = c = 1$, of inverse energy squared, inverse to those of $\phi$).

\subsection{Implications in Cosmology} \label{sec:cosmicimplications}

With the general result in hand, we now specialize to the standard cosmological model of a spatially flat FLRW spacetime $M = \mathbb{R} \times \Sigma$ on which the metric locally takes the form
\begin{equation} \label{eqn:flrw}
g = -dt^2 + a(t)^2 \left[ dx^2 + dy^2 + dz^2 \right],
\end{equation}
coupled to a scalar field $\phi$ through (\ref{eqn:einsteinkg}) (modulo terms for regular matter and radiation). It is well known that the homogeneity and isotropy of this model ensures that $\phi$ is constant on the spatial slices $\{t\} \times \Sigma \subset M$ for each fixed $t$, so that $\phi = \phi(t)$ is only a function of cosmological time \cite{bray2010}. In particular, the Klein Gordon equation takes the form of a damped oscillator equation
\begin{equation} \label{eqn:cosmologicalkg}
\ddot \phi + 3 H \dot \phi + m^2 \phi = 0,
\end{equation}
where $H = \frac{\dot a}{a}$ is the Hubble parameter. The mass parameter $m$ directly takes the role of the oscillator's (angular) frequency, while the damping term $3H \dot \phi$ leads the amplitude to decay (once oscillations begin, when $H \lesssim m$) proportionally to $a^{-3/2} = (1+z)^{3/2}$ as the universe expands \cite{huiwitten}.

Under the usual approximation of cosmological averaging for the purposes of understanding redshifts of distant sources, then, we expect that the difference $\phi(p_2) - \phi(p_1)$ in (\ref{eqn:redshift}) relevant to the time-varying observed redshift $\tilde z$ of a source at fixed comoving distance corresponding to a standard redshift $z$ has two distinct oscillating components: the oscillation of $\phi$ at observation (the point $p_2$) at frequency $m$ and the oscillation of $\phi$ at emission (the point $p_1$) at the redshifted frequency $\frac{m}{1+z}$. The latter frequency is shifted precisely by the standard cosmological factor $1+z = \frac{a(t_2)}{a(t_1)}$ because it is purely due to the universe's expansion, not geodesic parameterization-- the distance between two light pulses emitted by the source at subsequent crests of $\phi$ expands by this factor by the time they reach the observer. The logarithm of (\ref{eqn:redshift}),
\begin{equation} \label{eqn:lnredshift}
\ln(1+\tilde z) = \ln(1+z) + C \left[ \phi(p_2) - \phi(p_1) \right],
\end{equation}
indicates that these oscillatory frequencies $m$ and $\frac{m}{1+z}$ should appear directly in the quantity $\ln(1+ \tilde z)$, potentially making this signal easy to pick out via Fourier techniques applied to $\ln(1+\tilde z)$. Moreover, the amplitudes of these oscillations, while not set absolutely due to the unconstrained parameter $C$, should be correlated in a specific way due to the $a^{-3/2}$ decay of $\phi$-- more distant sources at fixed comoving distance should exhibit larger oscillations in a directly quantifiable manner. 

The above characteristics can be well-captured by modeling the repeated measuring, over laboratory time $t$, of the observed redshift $\tilde z(t)$ of an object at fixed comoving distance by making the identifications $\phi(p_2) \sim \sin(m t)$, where we've shifted $t = 0$ to eliminate any phase and suppressed the present-day amplitude, and $\phi(p_1) \sim (1+z)^{3/2} \sin(\frac{m t}{1+z} - \delta)$, where $\delta$ is a  phase shift arising due to the time delay between emission and observation, set by the precise distance to the source. As an order of magnitude estimate, $\delta \sim mD$, with $D$ the comoving distance, so that this shift is sensitive to variations in distance on the order of $\frac{2 \pi}{m}$. Inserting these identifications into (\ref{eqn:lnredshift}) yields the qualitative expectation
\begin{equation} \label{eqn:efflnredshift}
\ln \left( \frac{1 + \tilde z}{1 + z} \right) \propto \sin(mt) - (1+z)^{3/2} \sin \left( \frac{mt}{1+z} - \delta \right).
\end{equation}

In regards to the timescales of these oscillations, we expressing the mass parameter $m$ in units of $10^{-22}$ eV as $m_{22}$ and observe that
\begin{equation} \label{eqn:mvalue}
m = m_{22} \cdot 10^{-22} \text{ eV} \approx \frac{2 \pi m_{22}}{(1.3 \text{ yrs})}
\end{equation}
(recall we've set $\hbar = 1$), so that the frequency $m$ corresponds to an oscillatory period of about $\frac{1.3}{m_{22}}$ years. Since observational constraints largely point to $m_{22} \gtrsim 1$, we conclude that typical treatments of a cosmological scalar field as a viable primary dark matter candidate would lead to redshift oscillations in the theory developed here with period on the order of $\sim 1$ year or shorter as well as larger-amplitude oscillations at a redshifted period $1+z$ times longer. That these oscillations might occur on terrestrial timescales is a remarkable feature allowing the possibility of a comparatively simple means of detecting a signal from this instantiation of geometric scalar field dark matter.

Before turning to some preliminary analysis of redshift data, we reflect on how one would expect this signal to emerge in practice. We first observe that, though the oscillation amplitude discussed above should increase proportionally to $(1+z)^{3/2}$ as we look at more distant sources, this does not mean that we should expect exorbitantly large oscillations in the logarithm of the CMB temperature (some $1100^{3/2} \sim 3.6 \times 10^4$ times larger than any present oscillations), the most distant source we can observe, even over long timescales. This is because the CMB is not emitted at a fixed comoving distance, but rather at a fixed (range of) time, so that $\phi(p_1)$, the scalar field at emission (appropriately averaged over emission times according to the recombination visibility function), does not change as we repeatedly observe the CMB. 

A separate consideration arises for spatially extended sources, those larger than a few times $ \frac{2 \pi}{m} \sim \frac{1}{m_{22}}$ lyr. Light received from such sources at a given observation time would have been emitted over a range of emission times spanning several periods of the oscillation in $\phi(p_1)$, washing out this contribution to $\ln(1+\tilde z)$ (while perhaps broadening spectral peaks)-- in (\ref{eqn:efflnredshift}), this amounts to summing many different spectra with an effective continuum of values of $\delta$ that span a range much larger than $2\pi$, so that the upward and downward shifts due to the second term in (\ref{eqn:efflnredshift}) largely negate each other. Such extended sources, of course, are generally all that can be made out at even mildly high redshifts ($z \gtrsim 0.1$), likely nullifying the $(1+z)^{3/2}$ growth in practical observations. The only likely exceptions to this nullification are supernovae redshifts, though these are more difficult to monitor given their short lifespan. On the other hand, since the second term in (\ref{eqn:efflnredshift}) is expected to wash out for extended sources, oscillations in such sources would be entirely due to conditions at the point of observation-- that is, they should be {\it coherent} across all such sources, giving a powerful means of testing our theory.

In all cases the oscillation in $\phi(p_2)$, the scalar field at observation, should remain present, provided only that observations' exposure times are much shorter than $2\pi/m$. At small redshifts ($z \lesssim 0.1$), however, the amplitude of this oscillation (for a compact source) becomes sensitive to the source's precise distance due to the potential for both constructive and destructive interference between $\phi(p_1)$ and $\phi(p_2)$, or the two sinusoids in  (\ref{eqn:efflnredshift}). Indeed, standard trigonometric manipulations\footnote{$$A\sin(x) + B\sin(y) = (A+B)  \sin \left( \frac{x+y}{2} \right) \cos \left( \frac{x-y}{2} \right) + (A-B)\cos \left( \frac{x+y}{2} \right) \sin \left( \frac{x-y}{2} \right) $$} yield that in the limit $z << 1$, (\ref{eqn:efflnredshift}) effectively becomes
\begin{equation} \label{eqn:lowzefflnredshift}
\ln \left( \frac{1 + \tilde z}{1 + z} \right) \propto 2 \sin \left( \frac{z}{2+2z}mt + \frac{\delta}{2} \right) \cos \left( mt - \frac{\delta}{2} \right),
\end{equation}
wherein the cosine term gives the expected oscillation at frequency $m$, but modulated by the much more slowly-varying sine term setting the amplitude in a manner highly sensitive to the value of $\delta$. For $m_{22} \gtrsim 1$, this amplitude is sensitive to moving a source on the scale of a lightyear or less, meaning that amplitudes of the oscillations in low-redshift sources would be expected to be somewhat haphazardly distributed even at effectively fixed $z$. The factor of $2$ here means that the maximum amplitude is twice that expected from the $\sin(mt)$ term alone in (\ref{eqn:efflnredshift}), arising from potentially constructive interference. We comment that, at the larger end of the redshifts for which (\ref{eqn:lowzefflnredshift}) still gives a qualitatively correct picture ($z \sim 0.2-0.3$, though it would again be difficult to observe a compact source at such values), it becomes feasible that one might be able to observe both the frequency $m$ oscillations as well as their modulation on reasonable timescales.

From this investigation, then, we take away that in an aggregate view of many redshift variations across many sources, this model leads us to expect those at low redshifts to have oscillation amplitudes scattered between zero and a maximum value set by the parameter $C$ and the present-day amplitude of $\phi$, while higher redshift objects, generally being well beyond a lightyear in spatial extent, should exhibit oscillations which consistently attain about half this maximum amplitude. Moreover, oscillations of higher redshifts should be collectively coherent at frequency $m$. We remark that these conclusions are all made operating under the assumption that (\ref{eqn:redshift}) may be reasonably applied using the cosmologically averaged geometry of (\ref{eqn:flrw}). Though such assumptions have largely born out well in the standard cosmology, they merit further consideration in this modification, particularly given the large discrepancy generally expected between $\phi$ and its cosmological average at the points of emission and observation (being in galaxies) and the dependence of some features of this discussion on sub-lightyear scales. While an interesting problem, the resolution to this question is beyond the scope of this work, and we will simply assess whether the averaged predictions have any empirical support. Perhaps the best interpretation of this subsection's discussion is that it provides a heuristic motivation for seeking these signals rather than a robust prediction that they must occur precisely as described.

\section{Methods} \label{sec:methods}

To make a preliminary assessment as to whether the patterns discussed above are present in extant observational data, we make use of the Australian Dark Energy Survey's (OzDES) second data release \cite{lidman2020ozdes}, which catalogues high-quality redshift and spectroscopy data of some 30,000 sources up to redshift $z \sim 4$, with the highest priority sources being active transients, active galactic nuclei, and supernovae host galaxies. Each source in the catalogue was observed multiple times over the survey's duration from 2013 to 2019, annually between August and January, until the desired quality of redshift was obtainable from that source's stacked spectrum, an appropriately weighted average of all observations of the source of interest. Only this single, aggregate redshift was obtained and reported for each source in the catalogue, though the data release contained the individual spectra for each of their $\sim 375,000$ observations. The catalogue also contained data associated to observations of some 10,000 additional sources to which a redshift could not be confidently assigned, which we do not consider (in particular, we only considered sources with a redshift quality flag, assigned by OzDES, of at least 3).

As the patterns we seek to evaluate are in the time variation of the redshift $\tilde z(t)$ of individual objects, we need to assign a redshift to the individual observations' spectra rather than just each object's stacked spectrum. To have hope of extracting any meaningful representation of periodicity, we require many individual observations for each object we consider, so we restricted to those sources which were observed at least 30 separate times, reducing our data set to 1,457 sources with a total of 98,370 individual observations. For each source, we take the stacked spectrum's redshift reported by OzDES to represent the standard cosmological redshift $z$, as the averaging process should largely nullify the oscillations in (\ref{eqn:efflnredshift}), provided they occur over the data's 6 year timescale (we should obtain a null result otherwise). As we are ultimately interested in the relative quantity $\ln((1+\tilde z) / (1+z))$, we use each object's stacked spectrum as a baseline from which we ascertain a relative shift for each observation via template matching techniques.

\subsection{Identifying Redshift} \label{sect:identify}

Though discerning an optimal relative shift may seem like a straightforward task, some care must be taken to do this robustly. We first consider that if the unredshifted ``true" spectrum is $f(\lambda)$, then the stacked spectrum is expected to be $g(\lambda) := f((1+z) \lambda)$, and the observed spectrum is expected to be $h(\lambda) := f((1+\tilde z) \lambda)$. Describing the relative shift via $\alpha := \frac{1+\tilde z}{1+z}$, this means we expect $h(\lambda) = g(\alpha \lambda)$, and our computational task is to extract $\alpha$ from the data of $h$ and $g$ reported by OzDES. This will be made simpler with a logarithmic change of variables to turn the multiplicative shift by $\alpha$ into a linear shift by $\ln(\alpha)$. That is, defining $s := \ln(\lambda)$ and re-expressing the spectra as $\bar h(s) := h(\lambda(s)) = h(e^s)$ and 
$\bar g(s) := g(\lambda(s)) = g(e^s),$ the identity $h(\lambda) = g(\alpha \lambda)$ translates into

$$ \bar h (s) = h(e^s) = g(\alpha e^s) = g(e^{s + \ln(\alpha)}) = \bar g (s+\ln(\alpha)).$$
Our task is now to extract $\ln(\alpha)$, precisely the quantity in which our theory predicts oscillations, as the horizontal translation between the graphs of $\bar h$ and $\bar g$.

This is complicated in practice by random variations in noise, differing bulk atmospheric effects across the various observations, and the fact that such effects additionally mean that the spectra could not be consistently calibrated. Indeed, the OzDES documentation\footnote{https://docs.datacentral.org.au/ozdes/overview/dr2/} indicates: ``The spectra are not flux calibrated, not even in a relative sense. This is due to fibre positioning errors, chormatic [{\it sic}] aberrations from the 2dF corrector, and seeing." Our prescription for identifying the relative shift must therefore make the graphs of $\bar h$ and $\bar g$ most similar in an appropriate sense in light of these complications. A familiar tool for achieving this in general is the {\it cross-correlation} between $\bar h$ and $\bar g$:
\begin{equation} \label{eqn:corr}
(\bar h \star \bar g)(\tau) := \int_{\mathbb R} \bar h (s) \bar g(s+\tau)ds,
\end{equation}
a measure of the overlap between the graphs of $\bar h$ and the translational shift of $\bar g$ by $\tau$ to the left. Note that the bars here are part of the function notation, not complex conjugation-- all quantities are real. The value of $\tau$ which maximizes $\bar h \star \bar g$ would then be that which optimizes this overlap, providing a natural choice of $\ln(\alpha)$. A nice feature of (\ref{eqn:corr}) is that the maximizing value of $\tau$ is not affected by either vertical shifts or rescalings of either $\bar f$ or $\bar g$, so that concerns of callibration would be largely immaterial if we could actually work with this quantity. 

A practical complication to working with (\ref{eqn:corr}), however, is that one cannot observe the spectrum over all wavelengths-- the spectra with which we are working span about 3700\AA-8900\AA--, so the integral in (\ref{eqn:corr}) must be truncated to $ \int_a^b \bar h (s) \bar g(s+\tau)ds$ for some appropriate $a$ and $b$. Unfortunately, the adjustment imparted to the integral by adding a constant to $\bar h$ is now a function of $\tau$, so that the maximizing value of $\tau$ is no longer independent of vertical shifts. Moreover, the truncation can bias the maximal $\tau$ away from optimal alignment towards those shifts which move larger values of $\bar g$ into the integration range. Hence we must modify (\ref{eqn:corr}) beyond simply truncating. 

The bias due to the changing magnitude of $\bar g$ over $[a,b]$ can be countered by appropriately normalizing. Setting $\bar g_\tau (s) := \bar g(s+\tau)$ and noting that $ \int_a^b \bar h (s) \bar g(s+\tau)ds = \int_a^b \bar h (s) \bar g_\tau (s)ds$ is precisely the $L^2([a,b])$ inner product between $\bar h$ and $\bar g_\tau$, the most natural normalizing procedure would seem to be dividing by the $L^2$-norm. This is only strictly necessary for $\bar g_\tau$, as $\| \bar g_\tau \|$ depends on $\tau$ while $\| \bar h \|$ does not, but we also normalize $\bar h$ because it yields a universally meaningful quantity that can be used to compare the degree of correlation across different observations:
\begin{equation} \label{eqn:l2overlap}
C_N(\tau) = \frac{\langle \bar h, \bar g_\tau \rangle}{\|\bar h\| \| \bar g_\tau\|} = \frac{\int_a^b \bar h(s) \bar g_\tau(s) ds}{\sqrt{ \int_a^b (\bar h(s))^2 ds \cdot \int_a^b (\bar g_\tau (s))^2 ds }}.
\end{equation}
This is the {\it normalized cross-correlation}, a commonplace tool in the evaluation of redshifts \cite{kelson2003optimal, kurtz1998rvsao, simkin1974measurements, tonry1979survey}. The reasonableness of this quantity as a measurement of the similarity between $\bar h$ and $\bar g_\tau$ is supported by the Cauchy-Schwartz inequality for $L^2([a,b])$, which indicates that (\ref{eqn:l2overlap}) has magnitude at most $1$, and further that its magnitude is equal to $1$ if and only if $\bar g_\tau$ is (almost everywhere) a constant multiple of $\bar h$, which is almost precisely what we'd like to detect. 

\begin{figure}[t]
\centering
     \begin{subfigure}[b]{0.49\textwidth}
         \centering
         \includegraphics[width=0.75\textwidth]{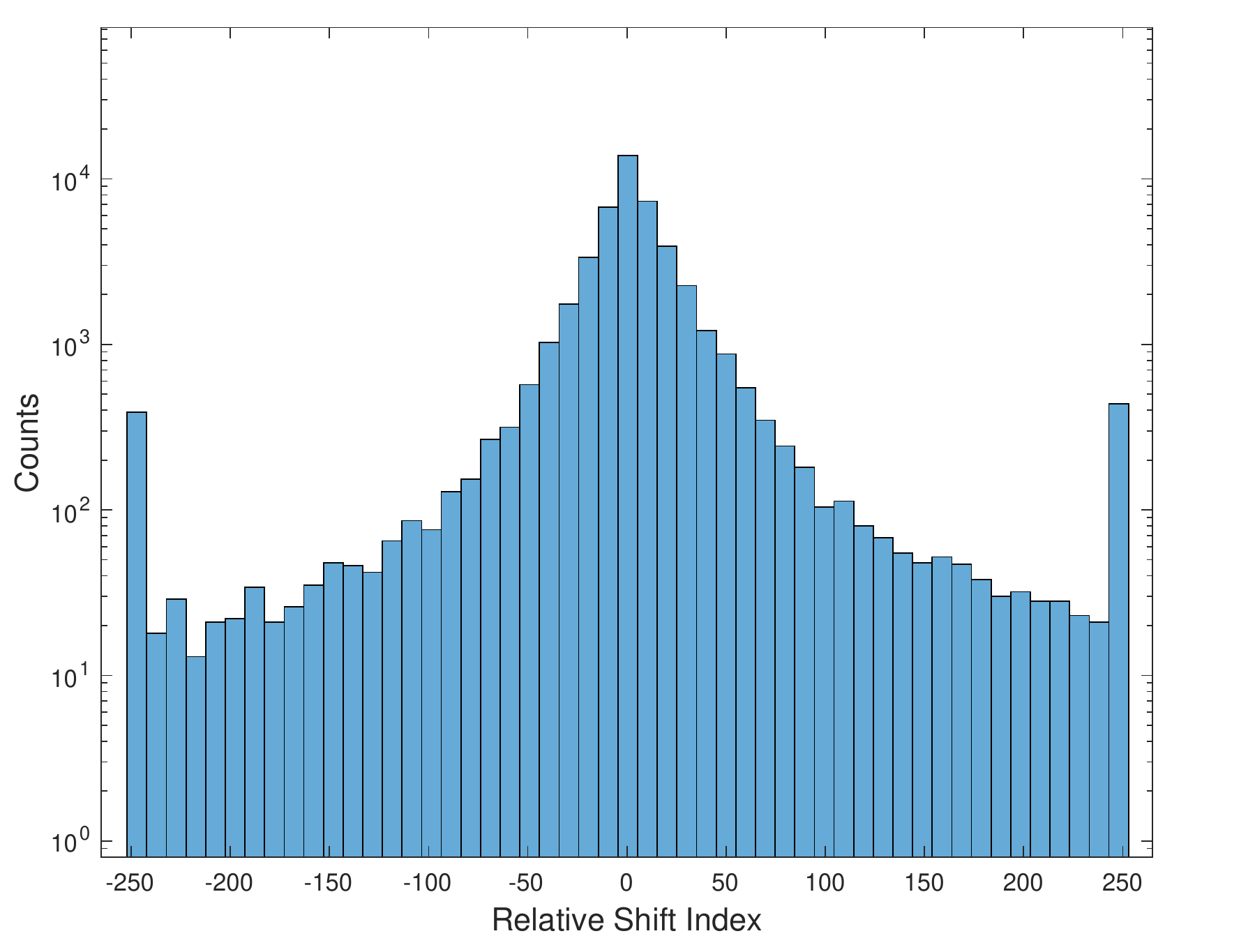}
     \end{subfigure}
     \hfill
     \begin{subfigure}[b]{0.49\textwidth}
         \centering
         \includegraphics[width=0.75\textwidth]{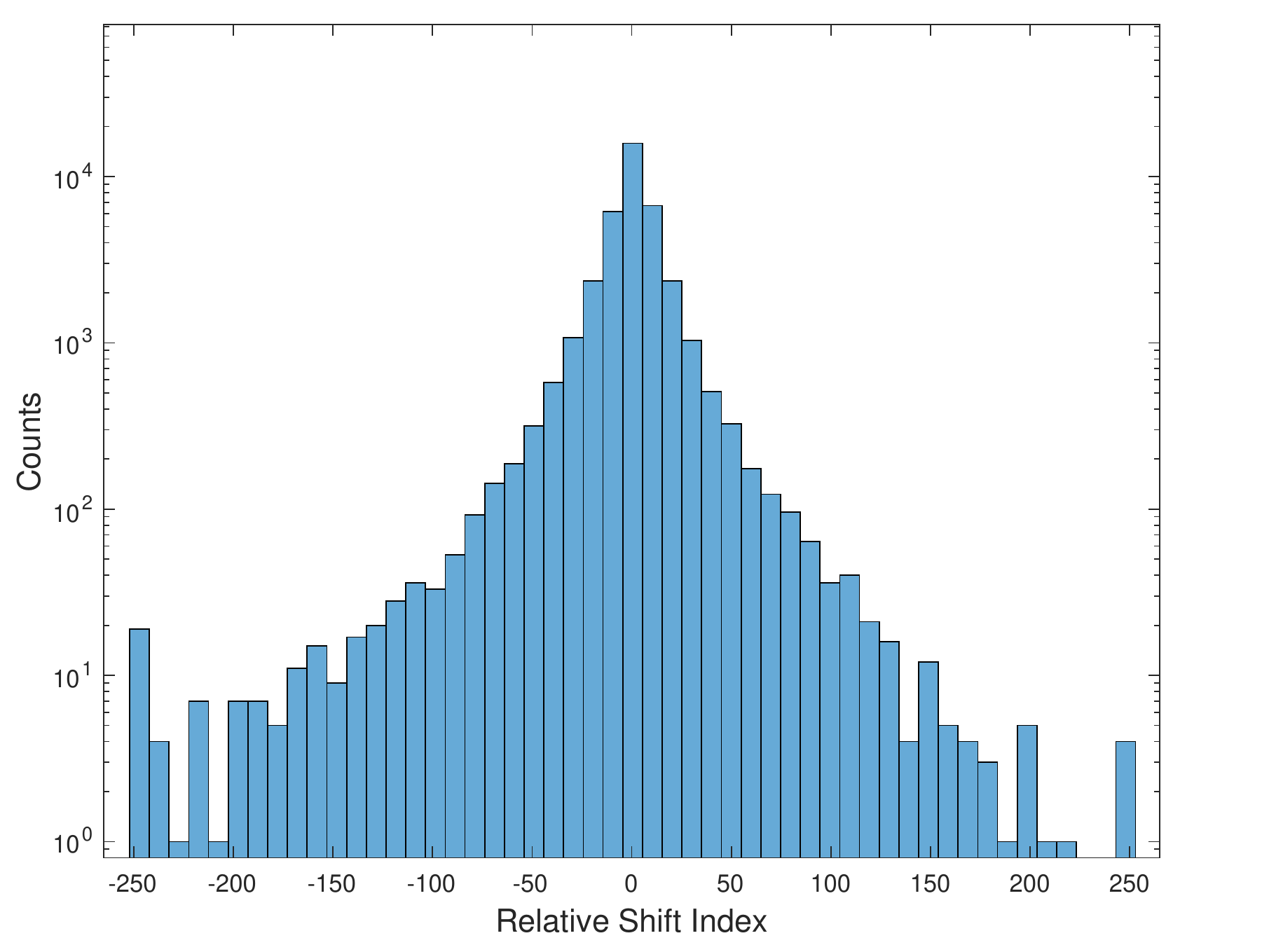}
     \end{subfigure}
\caption{ \footnotesize Incidence rates of relative shifts (after applying the cuts discussed in Section \ref{sec:program}), measured in terms of the $250$ steps between $0$ and the extreme shifts $\pm \tau_\text{max}$, in the cases that (left) constant averages and (right) large-scale Gaussian-weighted moving averages are removed. Note the logarithmic scaling of the vertical axis. The strategy on the right is subject to significantly less of the variation presumably induced by both incompatible calibrations and intrinsically varying continua between exposures.}
\label{fig:histograms}
\end{figure}

\begin{figure}[t]

\centering
     \begin{subfigure}[b]{0.49\textwidth}
         \centering
         \includegraphics[width=\textwidth]{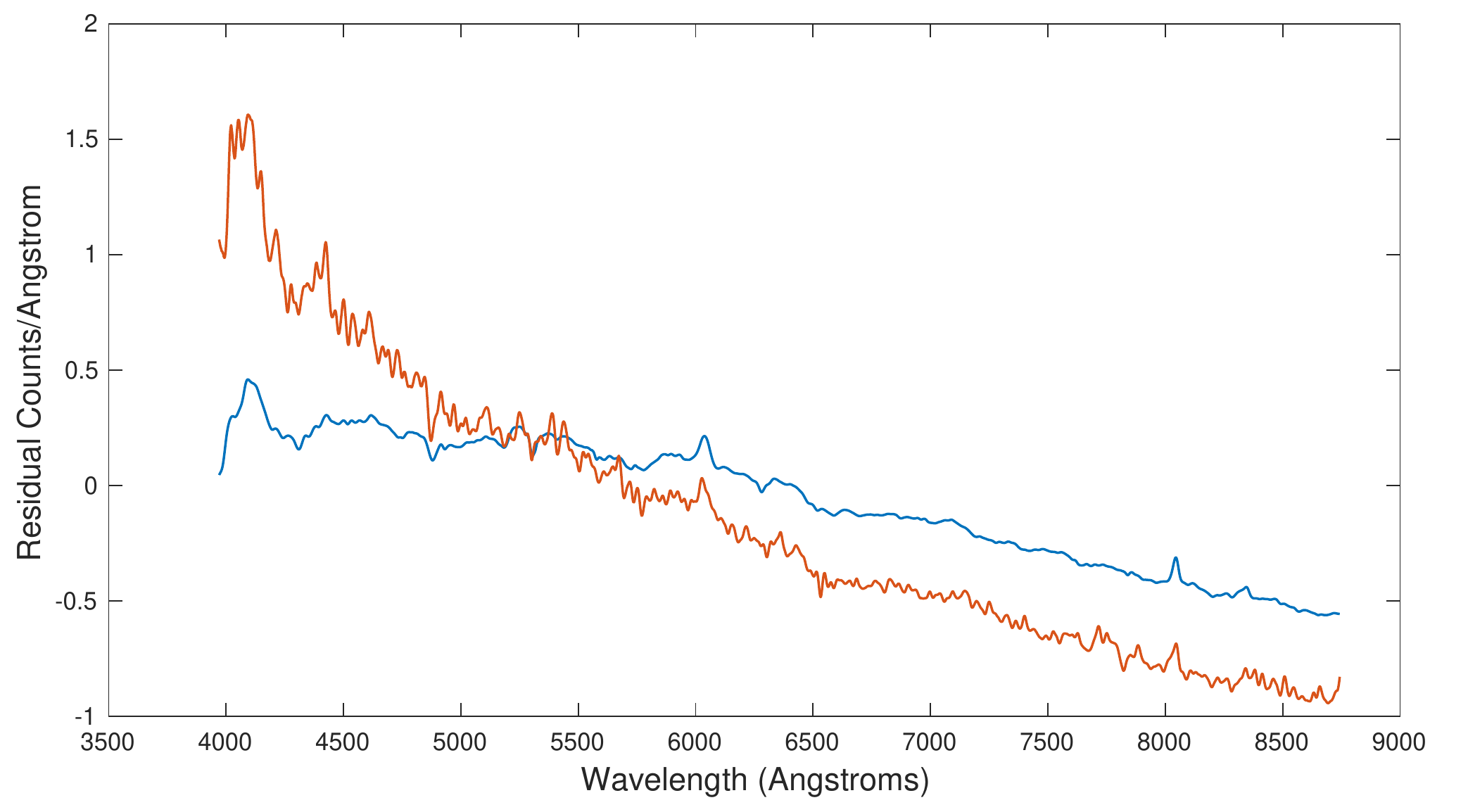}
     \end{subfigure}
     \hfill
     \begin{subfigure}[b]{0.49\textwidth}
         \centering
         \includegraphics[width=\textwidth]{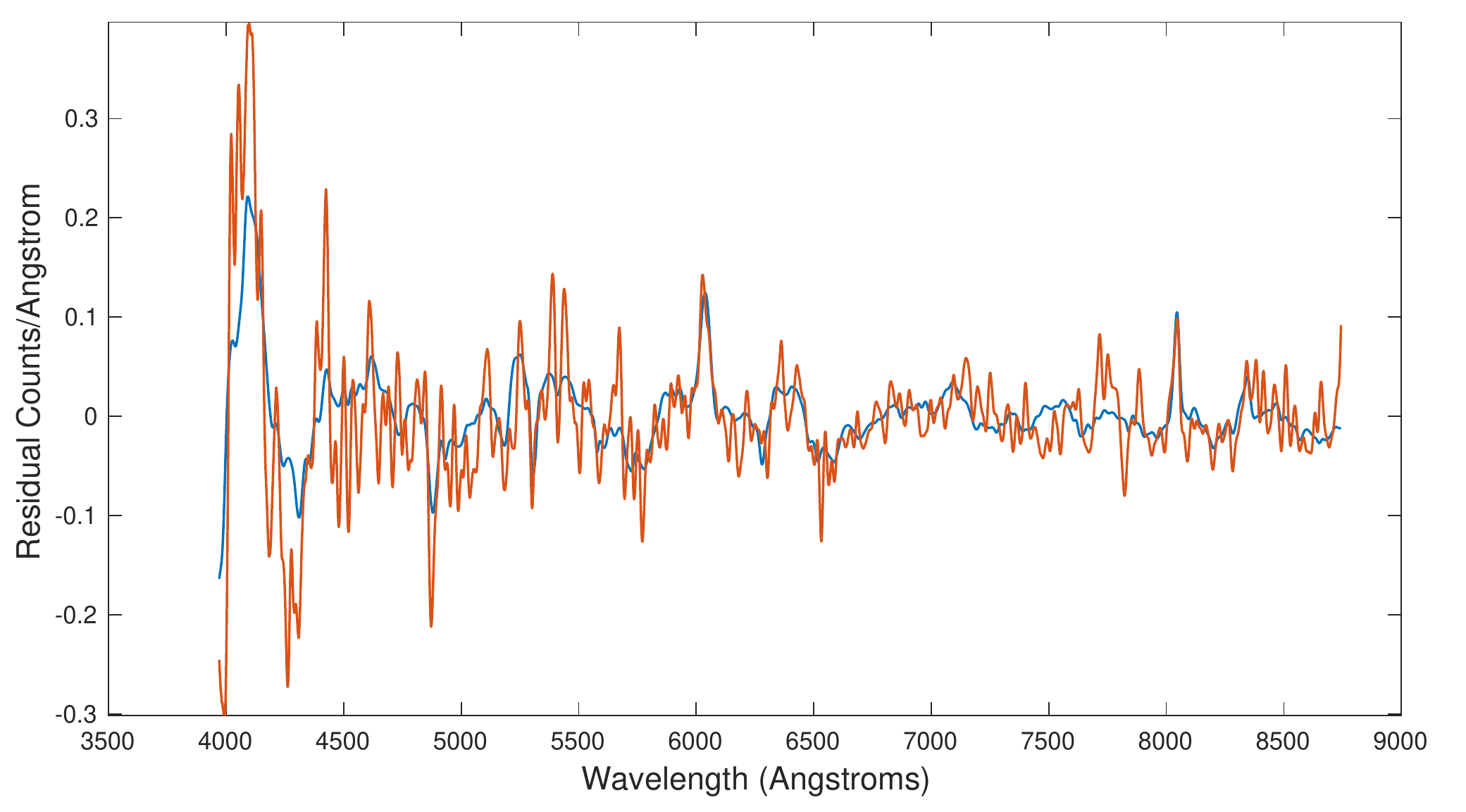}
     \end{subfigure}
     
     \begin{subfigure}[b]{0.49\textwidth}
         \centering
         \includegraphics[width=0.8\textwidth]{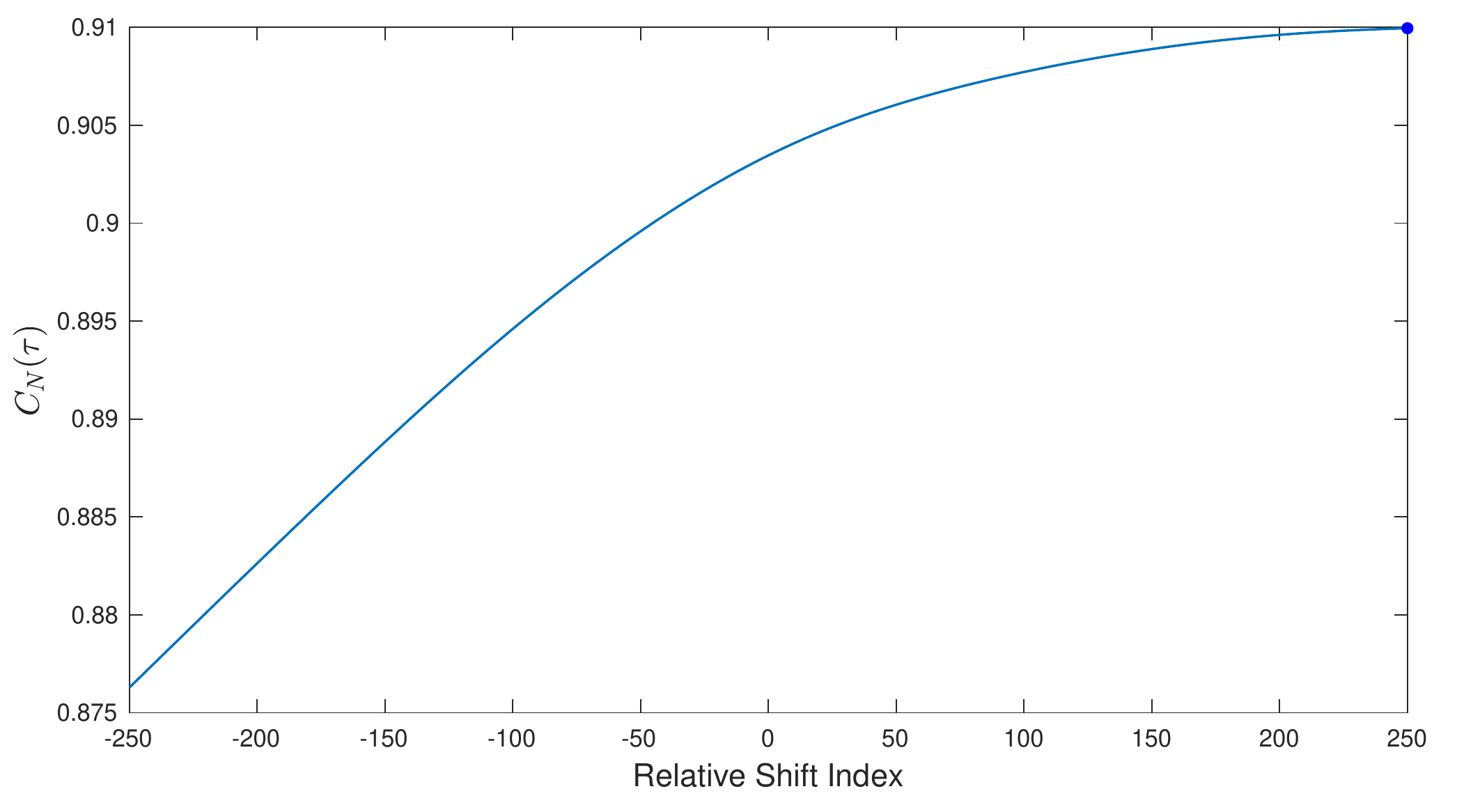}
     \end{subfigure}
     \hfill
     \begin{subfigure}[b]{0.49\textwidth}
         \centering
         \includegraphics[width=0.8\textwidth]{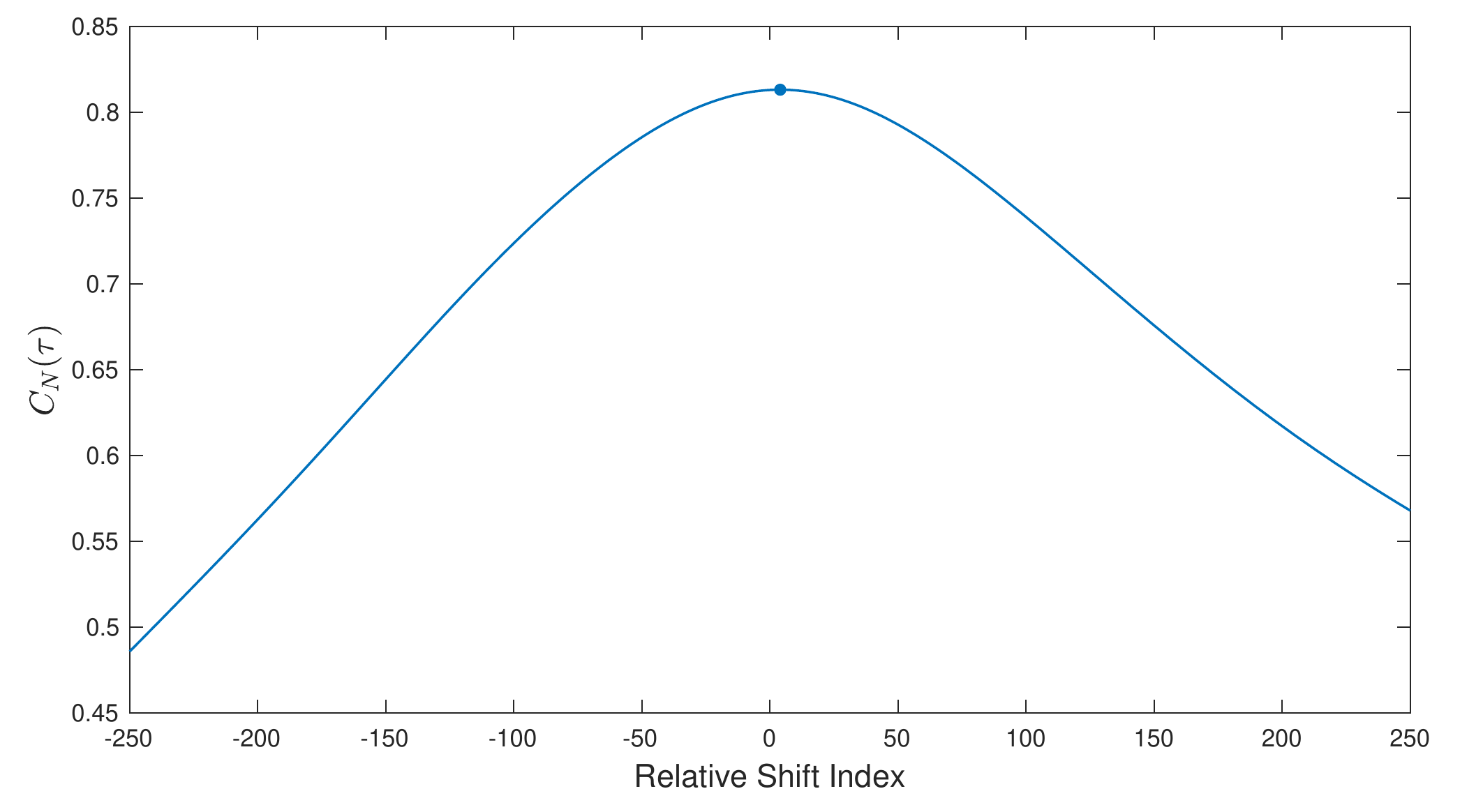}
     \end{subfigure}
\caption{ \footnotesize Top row: residuals of an observation (red) and stacked (blue) spectrum for a AGN source (named SVA1\_COADD-2971194281 by OzDES), upon subtraction of (left) constant averages and (right) large-scale Gaussian-weighted moving averages. On the left, one can see the change in calibration and/or continuum between the stacked and observation spectrum, while these are removed on the right, primarily leaving the closely aligned spectral features. 
\\ Bottom row: Plots of $C_N(\tau)$ for each of the pairs of spectra in the top row, with maxima emboldened. The maximum occurs at the largest probed shift $\tau_\text{max}$ on the left and very near the minimal shift on the right.}
\label{fig:residuals}
\end{figure}

The qualifier of ``almost" is used because the final concern to address is the potential need for a vertical shift to align the spectra. One approach to addressing concerns of this nature is to subtract away the average values of $\bar h$ and $\bar g_\tau$ before computing (\ref{eqn:l2overlap})-- that is, working with $\bar h - \frac{1}{b-a} \int_a^b \bar h(s)ds$ instead of $\bar h$, and similarly for $\bar g_\tau$. This is because the average behavior is very much susceptible to calibration concerns, and the spectral features by which redshifts are primarily identified are the variations on top of this average behavior anyway. These concerns are true of the average behavior more broadly than that contained in the average values, particularly since AGNs, which make up the bulk of our sources, exhibit varying spectral continua (this was a large part of what OzDES hoped to monitor, after all). Indeed, we've found in a number of cases that an appreciable bias can remain if we only subtract constant averages, leading to much wider variation in $\ln(\alpha)$ (see Figures \ref{fig:histograms} and \ref{fig:residuals}). Hence, we subtract a broader characterization of the average behavior, specified below, before evaluating (\ref{eqn:l2overlap}). Having done this, we identify $\ln(\alpha) = \ln ( (1+\tilde z)/(1+z))$ as the optimal value of $\tau$, that which maximizes $C_N(\tau)$.

\subsection{Programmatic Procedure} \label{sec:program}

The spectra catalogued by OzDES are reported on a wavelength range centered at $\lambda_c = 6295$\AA \; in $5000$ steps of width of about $1$\AA. To carry out the optimization discussed above, we maximized $C_N(\tau)$ among $501$ values of $\tau$ spanning the range between $\pm \tau_\text{max}$, where $\tau_\text{max}$ is defined so as to increase the central wavelength $\lambda_c$  by 50 steps, or $\tau_\text{max} \sim 0.0082$-- an initial evaluation allowed for shifting $\lambda_c$ by 200 steps, but no reasonably-confident shifts fell outside 50. The integration range utilized in computing (\ref{eqn:l2overlap}) was truncated by 230 steps at the lower end and 171 steps at the higher end so as to leave a buffer region from which data could be shifted into the range as $\tau$ is varied, yielding the window 3971\AA-8743\AA. All integrals needed in (\ref{eqn:l2overlap}) were computed via the trapezoid rule. All analysis was done with the 35-value Gaussian-weighted moving averages of the observed and stacked spectra $h$ and $g$ (using MATLAB's smoothdata function) to smooth out noise fluctuations occurring on the scale of several angstroms and mollify artifacts which yield large spikes in $h$, such as cosmic ray residuals \cite{lidman2020ozdes}. This smoothing is also how we characterized the average behavior to remove, identified as the Gaussian-weighted moving average over 1000 values.

Beyond smoothing, we applied a number of qualitative cuts to the data to address concerns surrounding poor data quality, reducing our effective data set. We did not consider observations for which the optimal correlation was poor, defined as the maximal $C_N(\tau)$ being less than $0.5$, as we took this to mean that spectral features were not strong enough to identify the redshift with confidence. Following \cite{lidman2020ozdes}, we further eliminated those observations which occurred during poor atmospheric conditions, evaluated via the catalogued zero points in the red and blue arms: we required both zero points to be greater than 30, with at least one greater than 31. The OzDES team also visually inspected most of their spectra, identifying a number of recurring spectral artifacts and recording them under the `QC' keyword in the FITS files-- we have ignored all observations which did not receive a flag of `ok'. Finally, we have removed those observations whose spectra exhibited exorbitant spikes, defined as occurring when the sum of the $25$ largest values of $|h|$ was more than 15\% of the sum of all values of $|h|$ (after smoothing and subtraction of average behavior), as such spikes exert undue influence on $C_N(\tau)$.

As with any such procedure, the schema outlined here may well still be subject to some pathologies, and it will not perfectly capture the appropriate shift in every case, but we maintain that it should be sufficiently robust to capture consistent trends across a wide array of data.

\section{Results} \label{sec:results}

\begin{figure}[t]
\centering
\includegraphics[width=10cm]{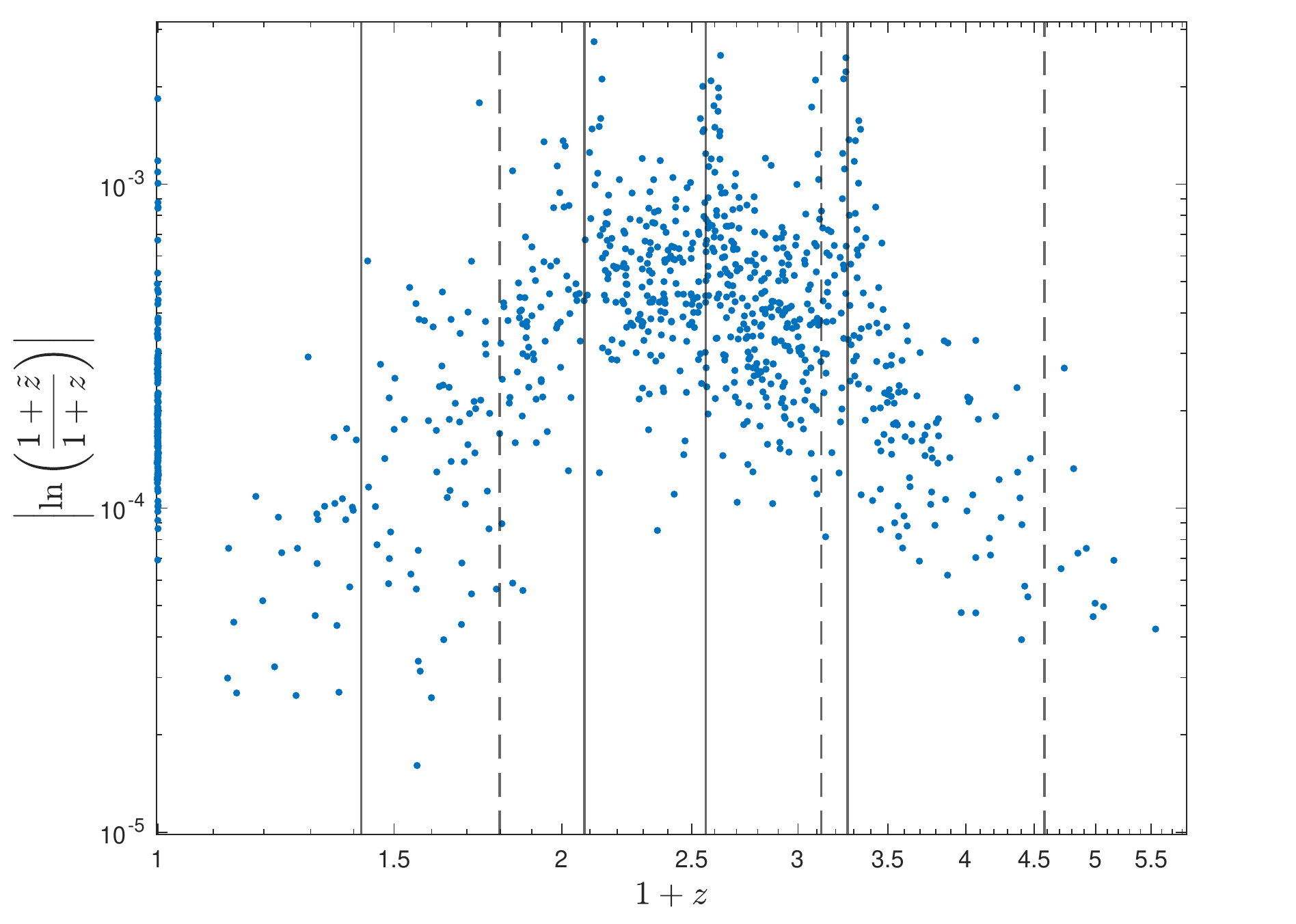}
\caption{ \footnotesize Mean redshift variation versus baseline redshift. Each point indicates the variation in the relative redshifts across individual exposure spectra for a single source. Also shown by vertical lines are the redshifts at which pertinent emission lines are shifted into (solid lines) and out of (dashed lines) the probed wavelength range. From left to right, these lines correspond to: MgII, H-$\beta$, CIII], CIV, MgII, Ly-$\alpha$, and CIII].}
\label{fig:redshiftvar}
\end{figure}

Applying all of the cuts discussed in the previous section leaves us with some 38,575 observations to which we've been able to assign a relative redshift with reasonable confidence, and these are associated to 902 sources which have at least ten admissible relative redshift values remaining. We first investigate the average magnitude of redshift deviations for each of these 902 sources, plotted against each source's baseline (stacked) redshift-- see Figure \ref{fig:redshiftvar}. To construct each point, then, we average the values of $|\ln(\alpha)|$ across the admissible observations associated to a given source. Irrespective of whether the data for the source in question is sufficiently fine to resolve periodic behavior in its redshift variations, the mean value of $|\ln(\alpha)|$ should, in aggregate across many sources, be a meaningful indicator of the amplitudes of oscillations in accordance with (\ref{eqn:efflnredshift}) that occur on timescales of a few years or less. Also overlain on this plot are vertical lines indicating the redshifts at which the most commonly strong emission lines enter or exit our integration interval.

We first note that the cluster of points around $z \approx 0$ in Figure \ref{fig:redshiftvar} corresponds to F stars used for throughput calibration, and these demonstrate a wide array of redshift variations with no readily apparent pattern. The primary feature among the remaining points, which generally correspond to AGNs and supernovae hosts, is the broad arch shape, indicating that redshifts in the range $1 \lesssim z \lesssim 2.5$ have relatively high variance, while both higher and lower values of $z$ yield significantly lower variance, decreasing as $z$ gets farther from this range, eventually reaching the rough scale of our minimum step size, $\frac{\tau_\text{max}}{250} \sim 3.3 \cdot 10^{-5}$. Also of note are the three apparent ``columns" exhibiting particularly high variance around $z \sim 1.1, 1.6,$ and $2.2$, which roughly line up with certain emission lines (CIII], CIV, MgII, and Ly-$\alpha$) transitioning into or out of the integration domain.

While the haphazard assortment of variances among F stars is not at odds with the general discussion of low $z$ in Section \ref{sec:cosmicimplications} (nor is it especially strong evidence in favor of it), the same cannot be said for the arch structure present throughout $z > 0.1$-- in particular, we do not see the redshift variance level off at and beyond $z \gtrsim 0.1$, as (\ref{eqn:efflnredshift}) led us to expect. Instead, the lines demarcating changes in spectral features paint the picture that these features are the dominant drivers of the variation in our identified shifts,  suggesting that this variation is uncertainty inherent to our technique rather than an indication of intrinsically varying redshifts. The variance predicted by (\ref{eqn:lnredshift}), then, apparently cannot contribute above the $\sim 10^{-4}$ level: this constrains the parameters $m$ and $C$ of the theory of Section \ref{sec:theory} to satisfy at least one of the order of magnitude constraints 
\begin{equation} \label{eqn:constraints}
m \lesssim 10^{-23} \text{ eV} \quad \text{\bf or} \quad m \gtrsim 10^{-18} \text{ eV} \quad \text{\bf or} \quad C \lesssim 10 \text{ eV}^{-2},
\end{equation}
corresponding to oscillations being either too slow to observe on the timescale of our data, too fast to be resolved given the typical instrument exposure time of $40$ minutes, or too small to be detectable over the fluctuations induced by spectral features. Note that the constraint for $C$ assumes $\Omega_\phi \sim 0.25$ in a cosmology with Hubble parameter $H \sim 70 \frac{\text{km}}{\text{s} \cdot \text{Mpc}}$ (each in order of magnitude). As discussed at the end of Section \ref{sec:cosmicimplications}, these constraints are subject to the caveat that they've assumed the legitimacy of cosmological averaging. More generally, the lack of signal detected here may instead translate to constraints on $C$ based on galactic dark matter densities rather than the cosmological average density, or constraints on $m$ based on the timescales of soliton or ``quasiparticle" periods in the Milky Way (generally much longer than $2\pi/m$) \cite{hamm2021scalar} rather than the cosmological oscillation timescale-- these are more convoluted threads to follow, and we will not attempt to do so in this work.

We now turn to whether there is any evidence for periodic behavior in $\ln(\alpha)$ as a function of time. Even if, as discussed above, the apparent variations in $\ln(\alpha)$ are largely due to limitations of our technique and the structure of the spectra, their behavior over time could still conceivably encode a preferred frequency extractable via Fourier techniques. No single source has enough observations for this to be done very meaningfully, but the conclusion of Section \ref{sec:cosmicimplications} that the oscillations of sources at $z \gtrsim 0.1$ should be coherent means that we may probe for an underlying frequency using data from all such sources at once.

\begin{figure}[b!]
\centering
     \begin{subfigure}[b]{0.49\textwidth}
         \centering
         \includegraphics[width=\textwidth]{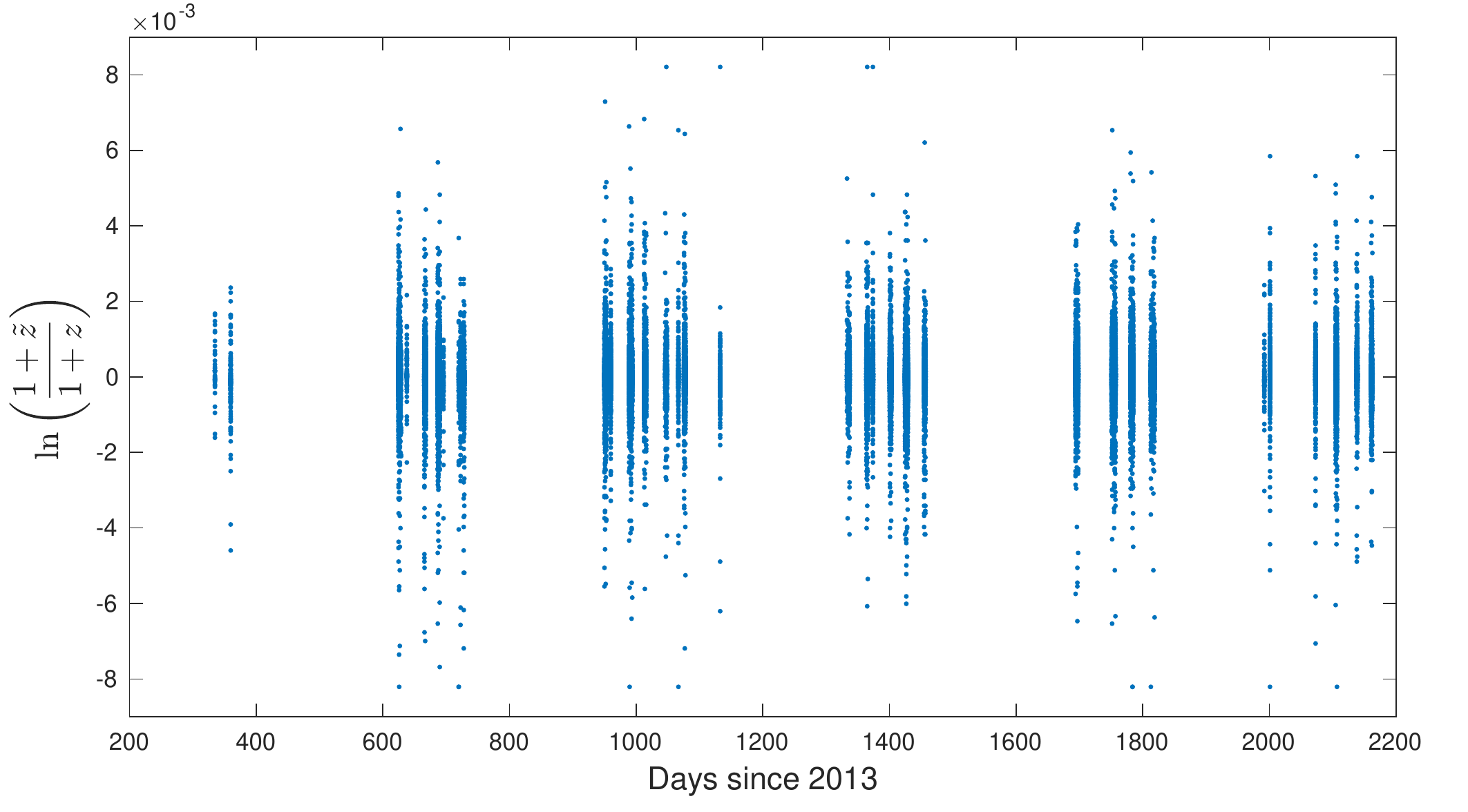}
     \end{subfigure}
     \hfill
     \begin{subfigure}[b]{0.49\textwidth}
         \centering
         \includegraphics[width=\textwidth]{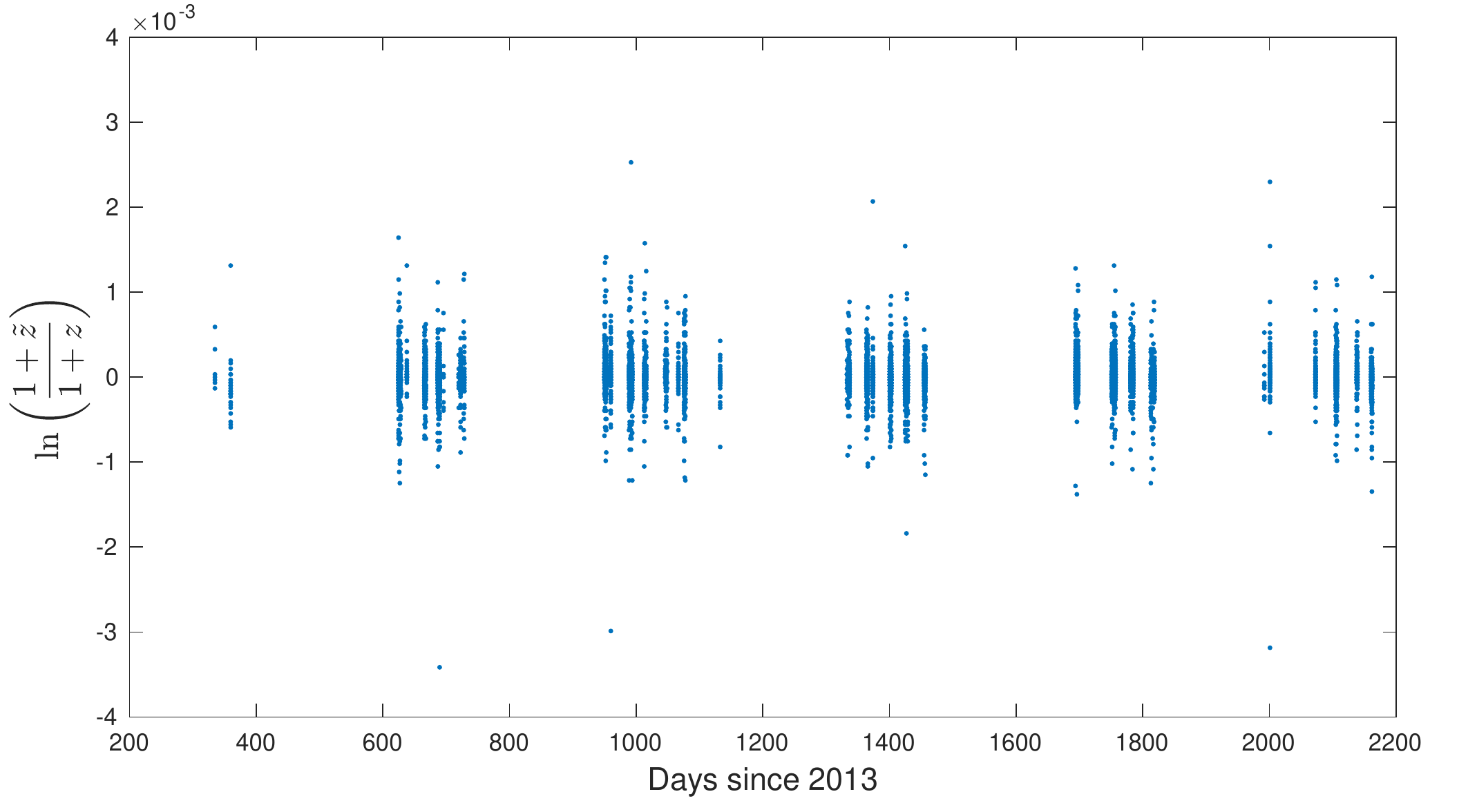}
     \end{subfigure}
\caption{ \footnotesize Relative redshifts plotted over time across all observations with (left) sources with $z > 0.1$ and (right) sources with either $0.1 < z < 0.7$ or $z > 2.5$.}
\label{fig:shiftvtime}
\end{figure}

\begin{figure}[t!]
\centering
     \begin{subfigure}[b]{0.49\textwidth}
         \centering
         \includegraphics[width=\textwidth]{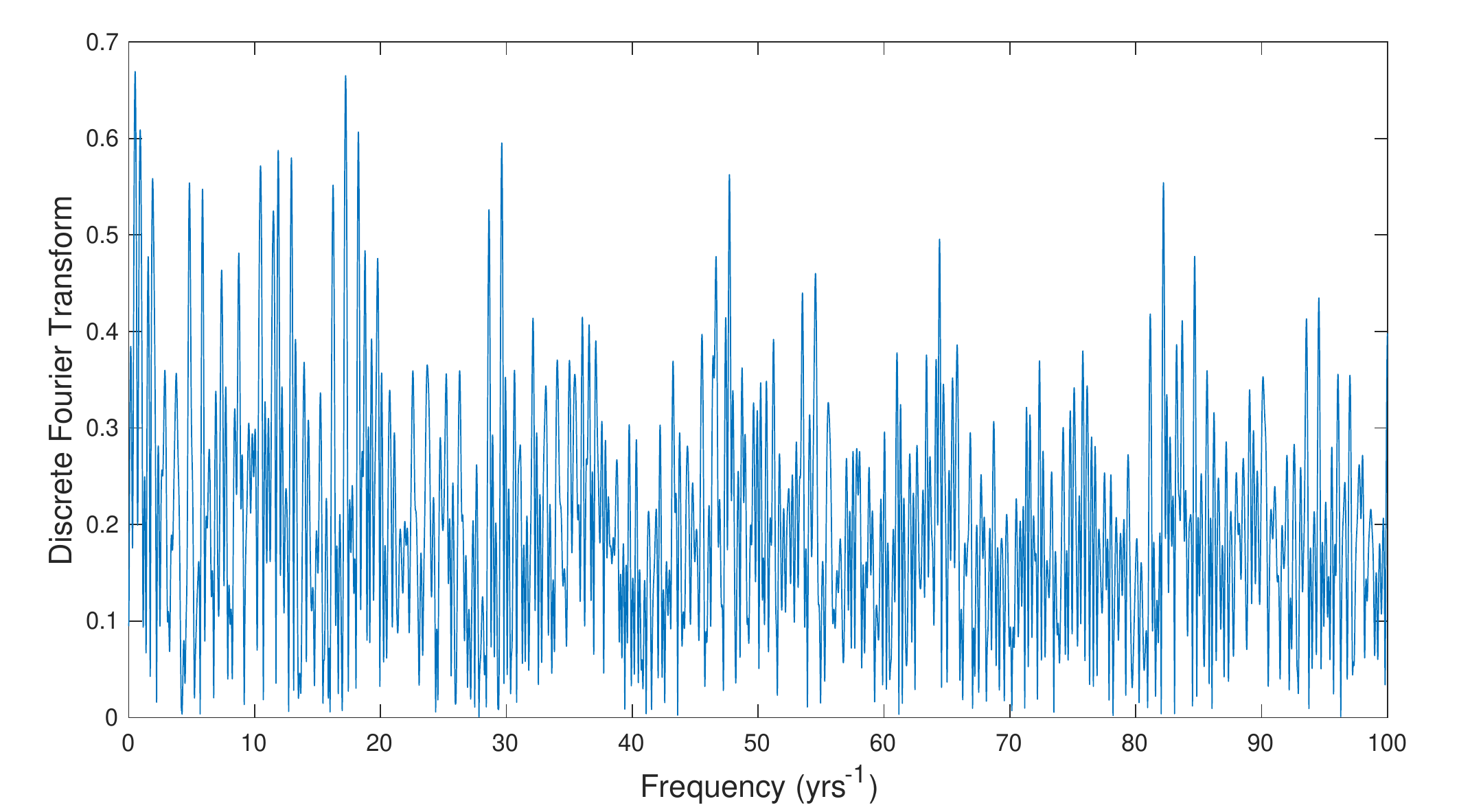}
     \end{subfigure}
     \hfill
     \begin{subfigure}[b]{0.49\textwidth}
         \centering
         \includegraphics[width=\textwidth]{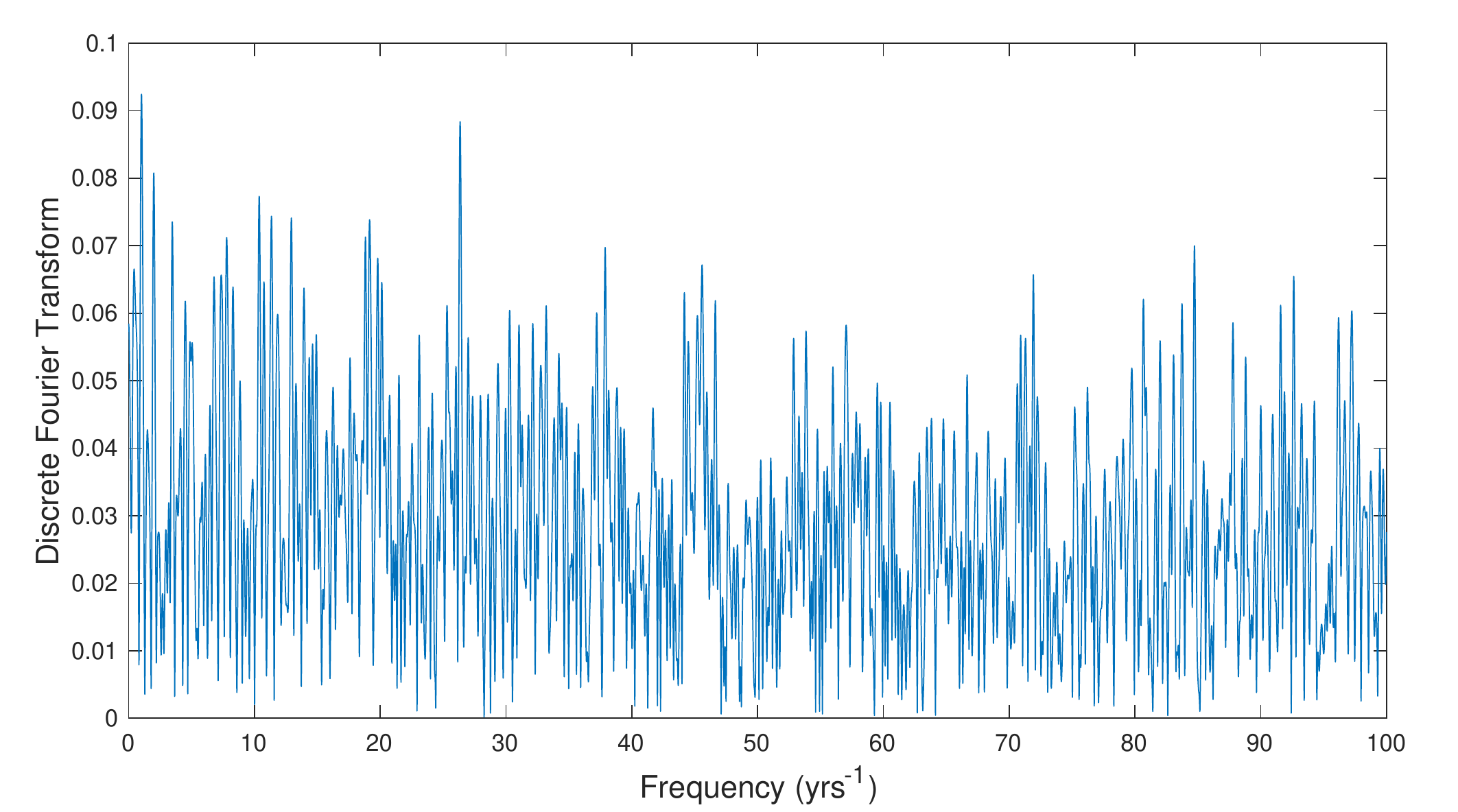}
     \end{subfigure}
\caption{ \footnotesize Magnitudes of the nonuniform discrete Fourier transforms of the two time series of Figure \ref{fig:shiftvtime}. The horizontal axis records linear frequency (as opposed to angular).}
\label{fig:dft}
\end{figure}

In the interest of broadly exploring the available data, we perform this probe with two different data sets: one comprised of all 33,727 observations at $z > 0.1$ to which we've assigned a relative redshift, the other comprised of the subset of 6922 observations which are further constrained to not lie in the range $0.7 < z < 2.5$. This latter restriction is informed by Figure \ref{fig:redshiftvar}, which leads us to expect that we should acquire a better signal to noise ratio by excluding these intermediate redshifts. Spectrally, these are the observations which have either the H-$\beta$ or the Ly-$\alpha$ emission peak squarely contained within their wavelength windows. In Figure \ref{fig:shiftvtime}, we plot the relative redshift measure $\ln(\alpha)$ in each of these data sets against the observation timestamps, measured in days since December 31st, 2013, and in Figure \ref{fig:dft} we plot the nonuniform discrete Fourier transforms of each of these up to a frequency of $100$ yrs$^{-1}$ (computed with MATLAB's nufft function).

The most notable features in Figure \ref{fig:shiftvtime} are the gaps, both between the yearly observation schedules and between the observation runs in clusters generally a few weeks apart. A zoomed in view would reveal further gaps between the nightly observations in each run, and on all of these scales points are scattered vertically into columns-- many points even exist at identical timestamps, as spectroscopy data was taken for many sources simultaneously in each exposure. Though the scatter is much less in the reduced data set on the right (note the vertical axis limits), these qualitative features are present in both. While some columns do appear vertically higher or lower than others at a glance, neither plot exhibits any glaringly obvious periodicity on the whole, though it is difficult to be definitive given the gaps.

Absent any visually obvious periodicity, we turn to the discrete Fourier transforms shown in Figure \ref{fig:dft}. In keeping with our take away from Figure \ref{fig:shiftvtime}, here we see that there are no peaks set strongly apart from the noise. This is especially so in the case of the larger, noisier data set of the left plot, where the two largest peaks at $f \approx 0.51$ yrs$^{-1}$ and $f \approx 17.23$ yrs$^{-1}$ are accompanied by several other peaks of similar height (though we observe that nearly all sizable peaks beyond $20$ yrs$^{-1}$ seem to be harmonics of the latter). In the smaller data set on the right, the two largest peaks at $f \approx 1.02$ yrs$^{-1}$ and $f \approx 26.35$ yrs$^{-1}$ are marginally more distinguished, but still not exceedingly so. Of course, the data's generally being taken at 1 year intervals means the peaks at $0.51$ and $1.02$ yrs$^{-1}$ are somewhat suspect, and observing $17.23$ yrs$^{-1} \approx \frac{0.99}{3 \text{ wks}}$ and $26.35$ yrs$^{-1} \approx \frac{1.01}{2 \text{ wks}}$ renders these frequencies suspicious as well. Such patterns continue beyond the plotted range (e.g., peaks appear at $f \approx 365$ yrs$^{-1}$ as well). In any event, this analysis weakly brings out some frequencies of potential interest, but a data set that's more complete in the time domain would be helpful to making definitive conclusions. Again, the absence of a strongly preferred frequency may either mean that $m$ is incompatible with oscillations on these timescales or that $C$ is sufficiently small that they are obfuscated by the noise, as in (\ref{eqn:constraints}).

\section{Conclusions}   \label{sec:conclusions}

In this work, we have developed a novel theoretical prediction made by a particular instantiation of the geometric model for scalar field dark matter broadly described by Bray \cite{bray2010}, and we have explored its implications for one of the most important cosmological observables, redshifts of distant sources. This pursuit has led us, through equations (\ref{eqn:redshift}) and (\ref{eqn:efflnredshift}), to several features of the theory which readily lend themselves to empirical verification (given the reasonability of cosmological averaging), including broadly coherent oscillations, at the frequency of the scalar field's mass parameter $m$, in the time evolution of redshifts of sources at mildly high baseline redshifts ($z \gtrsim 0.1$), as well as the dependence of such oscillations' amplitudes on baseline redshift across all values, depending on the spatial scale of the source of interest.

To investigate these predictions, we've drawn on the observations made by the Anglo-Australian Telescope for OzDES, which catalogued spectroscopy data for many thousands of sources observed several times each over the course of six years. By maximizing the cross-correlation (\ref{eqn:l2overlap}) with the source's stacked spectrum, we have associated to each observation of interest a shift $\alpha = \frac{1+\tilde z}{1+z}$ relative to the source's baseline redshift $z$ reported by OzDES. Comparing the behavior of $\alpha$ across the catalogue, in its dependence on both time and baseline redshift $z$, to the present theory's predictions, we have not found any compelling evidence that these predictions are born out in empirical data. While this result does not rule out the geometric model for dark matter under consideration, the absence of a signal at the levels of redshift variance probed here have led us to the tentative order of magnitude constraints of equation (\ref{eqn:constraints}) on the free parameters $m$ and $C$ of the theory. At the very least, the analysis culminating in Figure \ref{fig:redshiftvar} is the first to our knowledge establishing the consistent empirical stability of cosmological redshifts over timescales of several years, conservatively at the level of one part (of $1+z$) in a thousand (still some six orders of magnitude too coarse to probe the standard model's order of variation from (\ref{eqn:zdot})).

While the investigation carried through here has yielded null results, the relative ease of potentially obtaining a positive identification of dark matter through these methods means it remains of interest to probe the time and redshift evolution of $\alpha$ in both more sensitive and complete cosmological data sets. Perhaps the largest shortcoming of the data set utilized here was its discreteness, seen in the large gaps present in Figure \ref{fig:shiftvtime}-- the 33,727 observation spectra represented there were collected over only 92 nights, and the average individual source we considered only had admissible observations from 14 separate nights across the six years. A more continuous observation schedule would improve both the confidence in the amplitude of $\alpha$ and the ability of Fourier techniques to pick out an underlying frequency. Beyond this, the spatial compactness of supernovae means their oscillations may have larger amplitudes by up to an order of magnitude at higher redshifts (as discussed in Section \ref{sec:cosmicimplications}), so these can also provide a route to improvement. The Time-Domain Extra-Galactic Survey (TiDES) is an upcoming cosmological survey with the capacity to provided a more complete data set informing constraints on this theory with higher frequency observations and smaller seasonal gaps \cite{lidman2020ozdes, swann20194most}.

\section*{Acknowledgements}

The author would like to thank his advisor, Dr.\@ Hubert Bray, for supporting this work and engaging in helpful discussions, as well as Dr.\@ Michael Troxel for readily offering several constructive comments and pointing the author toward OzDES for a relevant data set. We also thank Dr.\@ Christopher Lidman for helping to identify the source of structure in Figure \ref{fig:redshiftvar}, and in particular indicating which emission lines were likely to be most pertinent.

\bibliographystyle{plain}
%{\footnotesize \bibliography{\string~/Documents/references.bib}}
{\footnotesize \bibliography{references.bib}}

\appendix

\section{Deriving the Theory} \label{app:theory}
Here we give a brief discussion of the action giving rise to the theory under consideration, encapsulated in equation (\ref{eqn:newchristoffel}). As alluded to in the text, a general connection $\nabla$ on a semi-Riemannian manifold $(M,g)$ can be entirely characterized by its {\it difference tensor} with respect to the Levi-Civita Connection $\overline \nabla$,

$$D(X,Y,Z) = \langle \nabla_X Y , Z \rangle - \langle \overline \nabla_X Y , Z \rangle,$$
so the variation of the appropriate action $S[g, \nabla]$ with respect to $\nabla$ can be carried out by varying $D$. The difference tensor can itself be understood in terms of two independent components associated to $\nabla$, the {\it metric compatibility tensor}
\begin{align*}
M(X,Y,Z) & = \langle \nabla_Z X, Y \rangle + \langle X, \nabla_Z Y \rangle - Z \left( \langle X,Y \rangle \right )
\\ & = D(Z,X,Y) + D(Z,Y,X)
\end{align*}
and the {\it torsion tensor}
\begin{align*}
T(X,Y,Z) & = \langle \nabla_X Y - \nabla_Y X - [X,Y], Z \rangle
\\ & = D(X,Y,Z) - D(Y,X,Z).
\end{align*}
Together, these entirely characterize $D$, and hence $\nabla$, according to
\begin{align} \label{eqn:difference1}
D(X,Y,Z) = \frac{1}{2} \big [ & T(X,Y,Z) - T(Y,Z,X) + T(Z,X,Y)   \nonumber
\\ - & M(X,Y,Z) + M(Y,Z,X) + M(Z,X,Y) \big ].
\end{align}
Varying $\nabla$, then, is equivalent to independently varying $M$ and $T$.

To naturally introduce a nontrivial connection, Bray \cite{bray2010} considered the axiom that the action $S[g,\nabla]$ be quadratic in the metric derivatives $g_{ij,k}$ as well as the connection coefficients $\Gamma_{ijk}$ and their derivatives $\Gamma_{ijk,l}$, extending the similar axiomatization of the Einstein-Hilbert action in terms of $g$ alone. Due to the obstruction that squares of derivatives of the form $\nabla D$ violate the axiom by including terms quadratic in metric second derivatives, he conjectured (\cite{bray2010}, Conjecture 1) that the most general means of introducing a squared derivative of $D$ in the action (so as to obtain nontrivial second order equations of motion for $D$) in keeping with the axiom was through terms of the form $|d \omega|^2$, where $\omega$ is the fully antisymmetric part of $D$, a $3$-form satisfying
\begin{equation}
\omega(X,Y,Z) = \frac{1}{6} \left[ T(X,Y,Z) + T(Y,Z,X) + T(Z,X,Y) \right].
\end{equation}
The remaining contribution of $D$ to $S[g, \nabla]$, then, is quadratic in $D$ itself.
%, and the various components of $D$ involved will only have nontrivial equations of motion if they can be %contracted with $\omega$. 
At this point, Bray restricts to the representative simplest case of $D = \omega$ to obtain
\begin{equation} \label{eqn:action0}
S[g, \nabla] = \int_U \left[ R - 2 \Lambda - c_1 |d \omega|^2 - c_2 |\omega|^2 \right] dV
\end{equation}
in the interest of most directly demonstrating the emergence of the Einstein-Klein-Gordon system (\ref{eqn:einsteinkg}). Though it is the simplest case and demonstrates the generic result of (\ref{eqn:einsteinkg}), this is but one choice of many in keeping with the conjecture, and different choices will have their own version of the connection relations (\ref{eqn:oldchristoffel}) and (\ref{eqn:newchristoffel}) which we're interested in. We treat a marginally more general case below.

Adopting the hypothesis of the main text that $\nabla$ should manifest physically in the determination of the geodesic trajectories of test particles, we are led to the physical expectation that $\nabla$ should be metric compatible (so $M(X,Y,Z) = 0$), as this is the only geometrically natural means of enforcing the special relativistic constraint that geodesic motion preserves timelike or null behavior. Under this expectation, (\ref{eqn:difference1}) implies that $\nabla$ is entirely characterized by the torsion tensor $T$. %Denoting by $\alpha$ the trace form of $T$ in the first and third slots, so in coordinates
%$$\alpha_j = T_{ij}^{\; \; \, i},$$
%this is the only piece of $T$ contractible with $\omega$ other than $\omega$ itself, so these will be the only nontrivial components of $D$. Hence, the most general scenario consistent with Bray's axiom and conjecture together with the physical constraint $M(X,Y,Z) = 0$ is characterized by the relation
The simplest extension of the previously considered case of a fully antisymmetric $T$ is to allow $T$ to have, in addition to its fully antisymmetric part, a nontrivial trace-- due to its general antisymmetry in the second and third slots, it can only have one--, which we describe in terms of the 1-form $\alpha$ given in coordinates by
$$\alpha_j = T_{ij}^{\; \; \, i}.$$
When $M = 0$ and $T$ is entirely characterized by this trace form and its antisymmetric part $2\omega$, we may write the difference tensor as
\begin{equation} \label{eqn:difference2}
 D(X,Y,Z) = \omega(X,Y,Z) + \frac{1}{3} \left[ \alpha(Y) \langle X, Z \rangle - \alpha(Z) \langle X , Y \rangle \right ],
\end{equation}
and the most general corresponding action becomes
\begin{equation} \label{eqn:action1}
S[g, \nabla] = \int_U \left[ R - 2 \Lambda - c_1 |d \omega|^2 - c_2 |\omega|^2 - c_3 |\alpha|^2 - c_4 \langle *\omega, \alpha \rangle \right] dV,
\end{equation}
where again $*$ denotes the hodge star operation and the constants $c_i$ are parameters of the theory.
For the purposes of carrying out the variation of $S$, it is convenient to recast the roles of $\omega$ and $\alpha$ in terms of the vector fields $w := (*\omega)^*$ and $v := \alpha^*$, where the raised asterisk is the metric dual turning covectors into vectors and vice versa, turning (\ref{eqn:action1}) into
\begin{equation} \label{eqn:action2}
S[g, \nabla] = \int_U \left[ R - 2 \Lambda + c_1 (\text{div } w)^2 + c_2 |w|^2 - c_3 |v|^2 + c_4 \langle w, v \rangle \right] dV. 
\end{equation}

The variation of $\nabla$ in this action is equivalent to varying each of $v$ and $w$ independently, so this will be our approach. The easier of these is $v$-- considering any one-parameter variation of $v$ given by $s \mapsto v(s)$ with $\dot v := \frac{d}{ds} \big |_{s = 0} v(s)$, that $v(0)$ (abbreviated to $v$) is at a critical point of $S$ for each $U$ requires

$$ 0 = \frac{d}{ds} \bigg |_{s = 0} S = \int_U  \langle -2c_3 v + c_4 w, \dot v \rangle dV$$
for every choice of variation, and hence for every $U$ and every possible variational vector field $\dot v$ (compactly supported in $U$). This requires the relation
\begin{equation} \label{eqn:vw}
2c_3 v = c_4 w
\end{equation}
to hold at a critical configuration of $v$ and $w$. The same procedure for varying $w$ yields
\begin{align*}
0 & = \int_U \left[ 2c_1(\text{div } w) ( \text{div } \dot w)  + \langle 2c_2 w + c_4 v, \dot w  \rangle \right] dV 
\\ & = \int_U \langle -2c_1 \nabla(\text{div } w) + 2c_2 w + c_4 v, \dot w \rangle dV,
\end{align*}
utilizing the divergence theorem and dispensing with the boundary term due to the variation's being compactly supported in $U$. Hence the critical configuration must also satisfy
\begin{equation}\label{eqn:divw}
2c_1 \nabla ( \text{div } w) = 2c_2 w + c_4 v = \left( 2c_2 + \frac{c_4^2}{2c_3} \right) w.
\end{equation}
Taking the divergence of both sides of this equation and defining $m^2 := \frac{1}{2c_1} \left( 2c_2 + \frac{c_4^2}{2c_3} \right) $ and $ \phi := \frac{\text{div } w}{m^2}$ leads us to the Klein Gordon equation,
$$\Box \phi = m^2 \phi.$$

Having finally identified $\phi$, we are now in a position to unravel our equations to obtain (\ref{eqn:newchristoffel}). Taking the metric dual of (\ref{eqn:divw}) yields
\begin{equation} \label{eqn:dphi}
d \phi = *\omega,
\end{equation} 
while further taking the hodge star gives
\begin{equation}  \label{eqn:omega}
\omega = *d\phi.
\end{equation} 
This identifies the first term on the righthand side of (\ref{eqn:difference2}) in terms of $\phi$, and we may similarly identify the latter terms by substituting (\ref{eqn:dphi}) into the metric dual of (\ref{eqn:vw}), obtaining
\begin{equation} \label{eqn:alpha}
\alpha = \frac{c_4}{2c_3} (*\omega) = 3C d\phi,
\end{equation}
where we've set $C := \frac{c_4}{6c_3}$. Putting (\ref{eqn:omega}) and (\ref{eqn:alpha}) into (\ref{eqn:difference2}), then, yields
$$ D(X,Y,Z) = (*d\phi)(X,Y,Z) + C \left[ d \phi (Y) \langle X,Z \rangle - d \phi (Z) \langle X,Y \rangle \right],$$
which is equivalent to (\ref{eqn:newchristoffel}).

The variation of our action (\ref{eqn:action2}) with respect to $g$ results in the Einstein equation with the scalar field source $\phi$ as in (\ref{eqn:einsteinkg})-- as this is not the core focus of this work, and as the procedure is essentially the same as that for the action (\ref{eqn:action0}), we again refer the interested reader to Bray \cite{bray2010}.

\end{document}